\documentclass{iaus}
\usepackage{graphicx}

\title[Dividing line between massive planets and brown-dwarf companions] 
{A possible dividing line between massive planets and brown-dwarf companions}

\author[J. Sahlmann et al.]   
{J.~Sahlmann, D.~S\'egransan, D.~Queloz, \and S.~Udry}
\affiliation{Observatoire de Gen\`eve, Universit\'e de Gen\`eve, 51 Chemin Des Maillettes, 1290 Sauverny, Switzerland. email: {\tt johannes.sahlmann@unige.ch}}

\pubyear{2010}
\volume{276}  
\pagerange{000--000}
\jname{The Astrophysics of Planetary Systems: Formation, Structure, and Dynamical Evolution}
\editors{A. Sozzetti, M.G. Lattanzi \& A.P. Boss, eds.}
\begin{document}

\maketitle

\begin{abstract}
Brown dwarfs are intermediate objects between planets and stars. The lower end of the brown-dwarf mass range overlaps with the one of massive planets and therefore the distinction between planets and brown-dwarf companions may require to trace the individual formation process. We present results on new potential brown-dwarf companions of Sun-like stars, which were discovered using CORALIE radial-velocity measurements. By combining the spectroscopic orbits and Hipparcos astrometric measurements, we have determined the orbit inclinations and therefore the companion masses for many of these systems. This has revealed a mass range between 25 and 45 Jupiter masses almost void of objects, suggesting a possible dividing line between massive planets and sub-stellar companions.
\keywords{brown dwarfs, stars: low-mass, binaries: spectroscopic, stars: statistics,  techniques: radial velocities, astrometry}
\end{abstract}

\firstsection 
\section{Introduction}
Radial-velocity (RV) studies show that close-in ($<\!10$ AU) brown-dwarf companions to Sun-like stars are rare compared to planets and stars (\cite{Marcy:2000pb}, \cite{Halbwachs:2003kx}, \cite{Grether:2006kx}). Precision-RV surveys have found $\sim\!50$ of these objects with $M_2 \sin i = 13-80\, M_J$, which we adopt as working-definition of the brown-dwarf mass range, regardless of the object's formation history, composition, and membership in a multiple system (\cite{Sozzetti:2010nx}, \cite[Sahlmann \etal\ 2010]{Sahlmann:2010kx}, henceforth SA10). Brown-dwarf (BD) companions have also been found in close orbit around M-dwarfs and in wide orbits ($>\!10$ AU) around stars, but these are not discussed in this contribution. There are several proposed scenarios for the formation of brown-dwarf  companions, but for an individual object we can assume that it either formed like a planet or like a high mass-ratio binary (\cite{Leconte:2009fk}). In particular, the discovery of transiting candidates (CoRoT-3b, CoRoT-15b, WASP-30b) in very close orbit around their F-type host stars challenge the current planet taxonomy.

\begin{table} \caption{BD candidates in the CORALIE survey}
\label{tab:BD} 
\begin{center}
\begin{tabular}{r r r r r  r r r r r} 
\hline %
Object & Ref. & $M_2 \sin i$ & $M_2$ & BD? & Object & Ref. & $M_2 \sin i$ & $M_2$ & BD?  \\  
            &    	 	   & ($M_J$)      & ($M_J$) & &            &    	 	   & ($M_J$)      & ($M_J$) &  \\  
\hline 
HD3277 & (1) & 64.7& 344.2& No  & HD112758 & (2) & 34.0& 210.0& No  \\ 
HD4747 &(1) & 46.1& $\cdots$&   &HD154697 &(1) & 71.1& 151.9& No  \\\ 
HD17289 &(1) & 48.9& 547.4& No  &HD162020 & (4) & 14.4& $\cdots$&  \\ 
HD18445 & (2) & 45.0& 175.0& No  &HD164427A &(1) & 48.0& 269.9& No \\ 
HD30501 &(1) & 62.3& 89.6& No & HD167665 &(1) & 50.6& $\cdots$& \\ 
HD38529 & (3) & 13.4& 17.6& Yes  &HD168443 & (5) & 18.1& $\cdots$&\\ 
HD43848 &(1) & 24.5& 101.8& No  &HD189310 &(1) & 25.6& $\cdots$& \\ 
HD52756 &(1) & 59.3& $\cdots$&   &HD202206 & (6) & 17.4& $\cdots$&\\ 
HD53680 &(1) & 54.7& 226.7& No  &HD211847 &(1) & 19.2& $\cdots$&  \\ 
HD74014 &(1) & 49.0& $\cdots$&   &HD217580 & (2)& 68.0& 170.0& No\\ 
HD89707 &(1) & 53.6& $\cdots$&  & & & & & \\ 
 \hline 
\end{tabular} 
\end{center}
\vspace{1mm}
 \scriptsize{ {\it References:} (1)  \cite{Sahlmann:2010kx}, (2) \cite{Zucker:2001ve}, (3) \cite{Benedict:2010ph}, (4) \cite{Udry2002}, (5) \cite{Wright:2009fe}, (6) \cite{Correia:2005bs}.}
\end{table}

\section{Search for brown-dwarf companions in the CORALIE survey}
\begin{figure}[bh]
\begin{center}
 \includegraphics[width= \linewidth, trim = 1.2cm 0cm 2.5cm 0cm, clip=true]{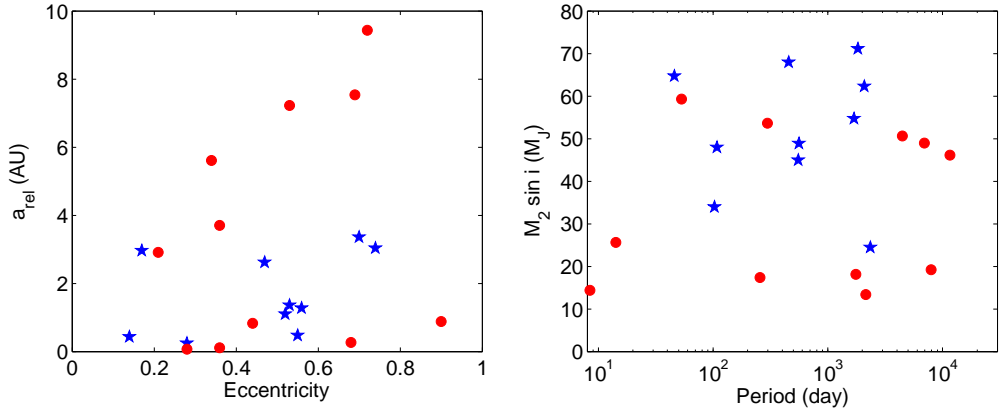} 
 \caption{Orbital parameters and $M_2 \sin i$ of the 21 potential brown-dwarf companions characterised with CORALIE. Blue asterisks show the M-dwarf companions and red dots indicate the remaining candidates and HD~38527. For display clarity, error bars are not shown. They can be found in the respective references given in Table~\ref{tab:BD}.}   \label{fig2}
\end{center}\end{figure}

On the basis of the well-defined and uniform CORALIE planet-search sample (\cite{Udry2000}), we have conducted a search for close BD companions of Sun-like stars (SA10). In combination with the astrometric measurements of Hipparcos, we were able to determine the orbit inclinations for many of these systems, which always rendered  M-dwarf companions. All 21 BD companion candidates in the CORALIE survey are listed in Table~\ref{tab:BD}. The orbital parameters of the potential BD companions, as derived from the RV solution of the respective reference papers, are shown in Fig.~\ref{fig2}. In relation to the sample size of 1600 stars, this yielded a frequency of 1.3 \% of candidate BD companions to Sun-like stars, assuming that all candidates have been discovered. Of these 21 candidates, ten are stellar companions with masses $M_2 > 80\,M_J$. Thus, less than 0.6~\% of Sun-like stars have close-in BD companions. One candidate was confirmed to a brown dwarf with HST astrometry (\cite[Benedict \etal\ 2010]{Benedict:2010ph}).

\begin{figure}
\begin{center}
 \includegraphics[width=0.7 \textwidth]{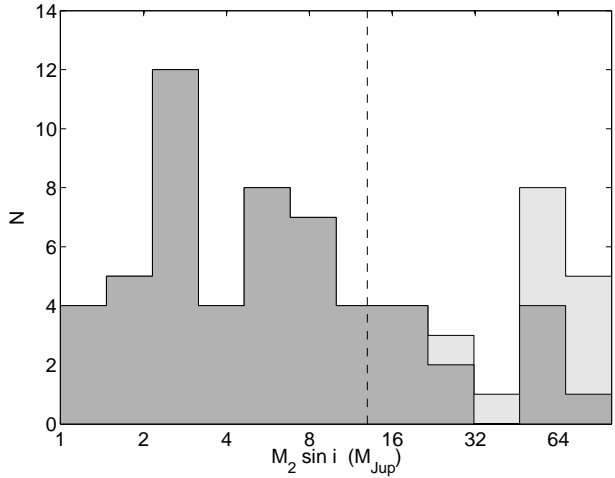} 
 \caption{Minimum mass histogram of sub-stellar companions characterised with CORALIE. The vertical dashed line indicates the $13\,M_J$ boundary. Light-grey areas show the companions for which SA10 found a mass higher than $80\,M_J$.}   \label{fig1}
\end{center}\end{figure}

Figure~\ref{fig1} shows the distribution of $M_2 \sin i$ for 85 sub-stellar companions characterised with CORALIE\footnote{The list of objects is taken from the series of CORALIE papers I-XVI (see \cite{Segransan:2010xr}), from SA10, and from Marmier et al. (\textit{in preparation}).}. Below $4\,M_J$, the sensitivity of the CORALIE survey is limited by the RV-measurment accuracy and timespan. At minimum masses higher than $4\,M_J$, the distribution shows approximately a linear decrease (in $\partial N/\partial \,\log M_2$) in the number of companions and extends into the brown-dwarf mass range. Above $45\,M_J$, a large population of companions appears, although it is drastically reduced by the astrometric analysis of SA10.  The $M_2 \sin i$ distribution of M-dwarf companions is shown in the light-grey histogram of Fig.~\ref{fig1}.

\section{Companion mass function}
\begin{figure}
\begin{center}
 \includegraphics[width= 0.7 \linewidth]{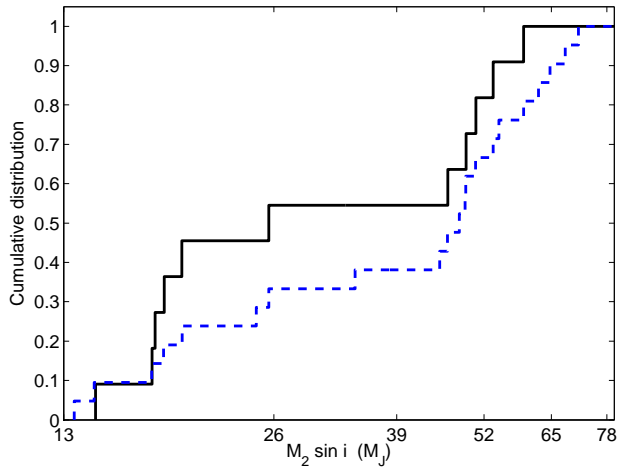} 
 \caption{Cumulative distribution of companions with $M_2 \sin i = 13-80\, M_J$ characterised with CORALIE (blue dashed line). The black solid line shows the distribution after removal of the M-dwarf companions (SA10).}   \label{fig4}
\end{center}\end{figure}

We show the cumulative distribution of minimum masses in Fig.~\ref{fig4}. The initial curve, including all 21 candidates, shows a steady increase of objects over the brown-dwarf mass range (dashed blue curve), though more than half of the objects have $M_2 \sin i > 45\, M_J$. After removal of the 10 stellar companions (solid black curve), the cumulative distribution exhibits a particular shape: it shows a steep rise in the $\sim\!13-25\, M_J$ region followed by a flat region spanning $\sim\!25-45\, M_J$ void of companions. At masses higher than $\sim\!45\, M_J$, the distribution rises again up to $\sim\!60\, M_J$. Above $\sim\!60\, M_J$, no companion is left.  

The distribution function's bimodal shape is incompatible with any monotonic companion mass function in the brown-dwarf and low-mass-star domain (approximately $13\, M_J - 0.6\, M_\odot$). Instead, the observed distribution can be explained by the detection of the high-mass tail of the planetary companions, which reaches into the brown-dwarf domain and contributes to the companions with $M_2 \sin i < 25\, M_J$, and the low-companion-mass tail of the binary star distribution with with $M_2 \sin i > 45\, M_J$. The interjacent mass-range is void of objects and represents a possible dividing line between massive planets and brown-dwarf companions obtained solely from observations. 

\section{Discussion}
The mass range of $25-45\, M_J$, which is void of companions, comprises the minimum of the companion mass function at $43^{+14}_{-23}\,M_J$ derived by \cite[Grether \& Lineweaver (2006)]{Grether:2006kx} for the 50~pc sample, which corresponds to the volume limit of the CORALIE survey. Hence, our result is in agreement with the conclusions of \cite[Grether \& Lineweaver (2006)]{Grether:2006kx}, which were based on extrapolation of the planet and binary distribution function into the brown-dwarf mass range. Still, the number of stars with brown-dwarf companions is limited, but both planet-search programmes (\cite{Diaz:2010, Santos:2010fk}) and more dedicated surveys (\cite{Lee:2010lr}) will provide us with many more candidates and will allow us to better trace the shape of the companion mass function. The GAIA astrometry mission (e.g. \cite{Lindegren:2010kx}) will detect and characterise a wealth of brown-dwarf companions and, not being affected by the inclination ambiguity, eventually allow us to identify their mass function. 

\end{document}